\documentclass[prd,aps,a4paper,preprint,preprintnumbers,nofootinbib]{revtex4-1}
\usepackage[a4paper, hdivide={1.91cm,,1.165cm}, vdivide={1.83cm,,3.3cm}]{geometry}

\usepackage{amsmath,amssymb}
\usepackage{graphicx,multirow}
\usepackage{color}
\usepackage{units}
\usepackage[hyperfootnotes=false]{hyperref}
\usepackage{soul}

\newcommand{\dd}{\mathrm{d}}
\newcommand{\ii}{\mathrm{i}}
\newcommand{\beq}{\begin{eqnarray}}
\newcommand{\eeq}{\end{eqnarray}}
\newcommand{\bea}{\begin{eqnarray}}
\newcommand{\eea}{\end{eqnarray}}

\begin{document}
\preprint{UCI-TR-2020-14}

\title{Understanding Q-Balls Beyond the Thin-Wall Limit}

\author{Julian~Heeck}
\email{heeck@virginia.edu}
\affiliation{Department of Physics, University of Virginia,
Charlottesville, Virginia 22904-4714, USA}

\author{Arvind~Rajaraman}
\email{arajaram@uci.edu}
\affiliation{Department of Physics and Astronomy, 
University of California, Irvine, CA 92697-4575, USA
}

\author{Rebecca~Riley}
\email{rriley1@uci.edu}
\affiliation{Department of Physics and Astronomy, 
University of California, Irvine, CA 92697-4575, USA
}

\author{Christopher~B.~Verhaaren}
\email{cverhaar@uci.edu}
\affiliation{Department of Physics and Astronomy, 
University of California, Irvine, CA 92697-4575, USA
}

\begin{abstract}
Complex scalar fields charged under a global $U(1)$ symmetry can admit non-topological soliton configurations called 
Q-balls which are stable against decay into individual particles or smaller Q-balls. These Q-balls are interesting 
objects within quantum field theory, but are also of phenomenological interest in several cosmological and 
astrophysical contexts. The Q-ball profiles are determined by a nonlinear differential equation, and so they
generally require solution by numerical methods. In this work,  we derive analytical approximations 
for the Q-ball profile in a polynomial potential and obtain simple expressions for the important Q-ball 
properties of charge, energy, and radius. These results improve significantly on the often-used thin-wall 
approximation and make it possible to describe Q-balls to excellent precision without having to solve the 
underlying differential equation.
\end{abstract}

\maketitle

\tableofcontents

\section{Introduction\label{s.intro}}

Under certain conditions, a scalar field theory
 admits the existence of localized non-topological soliton solutions of finite  energy.  
These so-called Q-balls are bound configurations of complex scalars $\phi$ that are stable against decay into individual 
particles or smaller Q-balls~\cite{rosen1968,Coleman:1985ki} (for a review, see Ref.~\cite{Nugaev:2019vru}). The complex 
scalars must carry a (global) $U(1)$ charge and require a special scalar potential~\cite{Lee:1991ax} or the inclusion of 
gravity as an attractive force~\cite{Colpi:1986ye}.
This simple setup can be modified in various ways. The most obvious is to make the $U(1)$ symmetry local, which 
leads to \emph{gauged} Q-balls~\cite{Lee:1988ag}. Another interesting extension is to include more than one 
scalar field in the soliton~\cite{Friedberg:1976me}. However, in this work we focus on the single-field, globally 
symmetric Q-ball.

Q-balls have been employed in many phenomenological and theoretical studies.
They have been analyzed as candidates for macroscopic dark matter of various 
types~\cite{Kusenko:1997si,Kusenko:2001vu,Ponton:2019hux,Bai:2019ogh}, 
including those similar to black holes and neutron stars. Many supersymmetric theories naturally predict 
Q-balls, the global $U(1)$ being identified with baryon or lepton number~\cite{Enqvist:1997si}. 
These Q-balls can play a role in dark matter~\cite{Kusenko:1997vp} or
baryogenesis as well as phase transitions in the early universe~\cite{Krylov:2013qe}.
They could lead to detectable gravitational wave signatures~\cite{Croon:2019rqu}.
Furthermore, Q-ball solutions are interesting in their own
right as a rare example of a stable non-topological soliton; their stability
is ensured by the conserved $U(1)$ charge as opposed to a topological charge.

Global Q-balls have been analyzed in a number of works. Their profiles are solutions to
a nonlinear differential equation that can only be solved analytically for some special 
potentials~\cite{Rosen:1969ay,Theodorakis:2000bz,MacKenzie:2001av,Gulamov:2013ema}. In general, the equations need to be solved 
numerically, which is straightforward but time consuming and
typically difficult when the size of the Q-ball is large.
Since the Q-ball ground state is the minimal-energy configuration for a fixed (large) charge $Q$, one can also minimize the 
energy functional with respect to a set of test functions in order to obtain analytic 
approximations~\cite{Ioannidou:2003ev,Ioannidou:2004vr}.

In this paper, we formulate an analytical approach for solving the
Q-ball equations and finding the profile in arbitrary sextic scalar potentials. As 
stable quartic potentials cannot produce 
Q-ball configurations, the scalar potential  
must have nonrenormalizable terms. In the low-energy effective theory, the leading such term respecting the symmetries would be sextic, $|\phi|^6$.
Consequently, by 
analyzing the general sextic potential we expect to encapsulate the leading dynamics relevant to most Q-ball systems of 
phenomenological interest.\footnote{Full models such as supersymmetic extensions unavoidably generate additional (potentially non-polynomial) couplings~\cite{Enqvist:1997si} that require dedicated analyses should they be sizable.  The methodology described here should be useful for any potential, though.} However, unlike some more specific potentials, the sextic has no known exact Q-ball solution. We 
show that while complete exact solutions remain elusive, very accurate analytical results can be derived near the 
large Q-ball limit.

When the Q-balls are large---that is, when their characteristic size is much larger than the mass term in the scalar 
potential---their defining equations can be dramatically simplified by 
considering the Q-ball as being a large 
object with  a small surface region \cite{Paccetti:2001uh,Tsumagari:2008bv}. 
Here we show that the profile near
the surface of a large Q-ball can in fact
be solved for exactly. This allows us to
find the full profile of the Q-ball in this limit. This novel result improves significantly, both qualitatively and quantitatively, upon the previously used thin-wall approximation. Our analytical profile and the derived Q-ball quantities (charge, energy, and radius) are in excellent agreement with numerical results in the large Q-ball limit. What is more, our 
results even describe smaller  stable Q-balls with $\mathcal{O}(10\%)$ accuracy, which can in principle be  
improved further.

Our results are also quite general in scope. By making a judicious choice of parametrization, we are able to reduce much 
of the Q-ball system to dependence on a single dimensionless parameter.  
We also find that the characteristics of the Q-ball are more naturally 
expressed as functions of the Q-ball radius, rather than the potential parameter. Thus, one of our primary results is 
determining how the Q-ball radius and the potential parameter relate to one another. We show explicitly how the difference 
between our analytic formulae and the numerical results can be reduced by improving the accuracy of this relationship.

We build up to these results by first, in Sec.~\ref{sec.review}, briefly reviewing Q-balls in order to establish the necessary 
notation and vocabulary. The thin-wall analysis introduced by Coleman is also presented. The analytical review is followed, in Sec.~\ref{sec.numerics}, by an outline of different 
numerical methods for obtaining Q-ball solutions. This includes a novel approach in which the radial coordinate is 
compactified. In Sec.~\ref{sec.improved_thinwall}, we derive new analytic approximations for Q-balls in and beyond the thin-wall limit. Our main results include simple formulae for the radius, charge, and energy of the Q-balls as well as an accurate ansatz for the scalar profile. These results
are then compared to the exact numerical results in Sec.~\ref{sec.comparison}.
We conclude in Sec.~\ref{sec.conclusion}, and we include some technical details on our final Q-ball profile, that are not 
necessary in order to understand the main text, in Appendix~\ref{app.matching}. 
 
\section{ Review of Q-Balls}
\label{sec.review}

Our starting point is the Lagrange density $\mathcal{L}$ for a complex scalar $\phi$ with potential $U(|\phi|)$,
\begin{align}
\mathcal{L} =\left|\partial_\mu\phi \right|^2-U(|\phi|) \, .
\end{align}
We choose the vacuum to be at $|\phi|=0$, and 
 the potential to be zero in the vacuum, so $U(0)=0$. 
This Lagrange density then exhibits a global $U(1)$ symmetry $\phi \to e^{\ii \theta}\phi$ associated with conserved $\phi$ 
number. 
 To ensure that the vacuum is a stable minimum, we require
\begin{align}
\frac{\dd U}{\dd|\phi|}=0\,, \qquad m_\phi^2\equiv\left.\frac{\dd^2U}{\dd\phi\, \dd\phi^\ast}\right|_{\phi=0}>0 \,,
\end{align}
where $m_\phi$ is the mass of $\phi$.
Coleman~\cite{Coleman:1985ki} showed that nontopological solitons, which he called Q-balls, can exist
in this theory if $U(|\phi|)/|\phi|^2$ has a minimum at $|\phi|=\phi_0/\sqrt{2}>0$ such that
\beq
0\leq \frac{\sqrt{2U(\phi_0/\sqrt{2})}}{\phi_0}\equiv\omega_0<m_\phi\,.
\label{e.Omega0}
\eeq
These Q-balls are spherical solutions to the resulting equations of motion which only depend on time through the phase 
of $\phi$, that is 
\begin{equation}
\phi (x) =\frac{\phi_0}{\sqrt{2}}f(r)e^{\ii \omega t}\,,
\end{equation}
for some constant $\omega$, thus evading Derrick's theorem~\cite{Derrick:1964ww}. Here, $f(r)$ is a dimensionless 
function of the radius $r\in [0,\infty)$ whose form is governed by the simpler Lagrangian
\begin{equation}
L=\int \dd^3 \vec{x}\, \mathcal{L} = 4\pi\phi_0^2\int \dd r\,r^2\left[ -\frac12f^{\prime 2}+
\frac12f^2\omega^2-U(f)/\phi_0^2\right],\label{e.Lag}
\end{equation}
where primes denote a derivative with respect to $r$. The resulting differential equation for $f$ is
\beq
\left(r^2f' \right)'=\frac{r^2}{\phi_0^2}\frac{\dd U}{\dd f}-r^2\omega^2f \,.
\eeq
Localized solutions to this equation are called Q-balls, and have a conserved global charge $Q$ and mass or energy $E$ 
given by the following integrals:
\begin{align}
Q &\equiv \ii\int \dd^4 x\, \left(\phi^\ast\partial^0\phi-\phi\partial^0\phi^\ast \right) \nonumber\\
& =4\pi\omega\phi_0^2\int \dd r\,r^2f^2\,,\label{e.Qdef}\\
E &= 4\pi\phi_0^2\int \dd r\,r^2\left[ \frac12f^{\prime 2}+\frac12f^2\omega^2+U(f)/\phi_0^2\right] \nonumber\\
&=\omega Q+4\pi\phi_0^2\int \dd r\,r^2\left[ \frac12f^{\prime 2}-\frac12f^2\omega^2+U(f)/\phi_0^2\right] .
\label{e.Edef}
\end{align}
Without loss of generality, we choose $\omega>0$, which implies $Q>0$.
The following relationship between $E$ and $Q$ holds for all Q-balls~\cite{Friedberg:1976me}:
\begin{align}
\frac{\dd E}{\dd \omega} &= \omega\frac{\dd Q}{\dd \omega}+Q+4\pi\phi_0^2\int \dd r\,r^2\left[f'\frac{\dd f'}{\dd\omega}
-\omega^2f\frac{\dd f}{\dd \omega}-f^2\omega+\frac{1}{\phi_0^2}\frac{\dd U}{\dd \omega} \right]\nonumber\\
&= \omega\frac{\dd Q}{\dd \omega}+Q-Q+4\pi\phi_0^2\int \dd r\,r^2\left[-\frac{\dd f}{\dd \omega}\left(\frac{1}{\phi_0^2}
\frac{\dd U}{\dd f}-\omega^2f \right)-\omega^2f\frac{\dd f}{\dd \omega} +\frac{1}{\phi_0^2}\frac{\dd U}{\dd \omega} \right]
\nonumber\\
&= \omega\frac{\dd Q}{\dd \omega}\,,
\label{e.dEdw}
\end{align}
where we have integrated one term by parts and used the equation of motion in the second line. For 
$\dd Q/\dd \omega\neq 0$, this implies $\dd E/\dd Q =\omega$ and allows us to interpret $\omega$ as a 
chemical potential. That is, $\omega$ determines how the energy changes when a particle of charge $Q$ is 
added or removed from the Q-ball. Clearly, when $\dd Q/\dd \omega>0$, it is energetically favorable for a 
given Q-ball to shed particle quanta to lower the energy, and indeed this condition is sometimes used to 
determine if a given Q-ball is stable. In the opposite case, $\dd Q/\dd \omega<0$, it is energetically 
favorable for the Q-ball to add particles.

The energy integral [Eq.~\eqref{e.Edef}] can be simplified further by employing an identity pointed out 
in Ref.~\cite{Lee:1988ag}. One may rescale the radial coordinate 
$r\to\chi\, r$~\cite{Derrick:1964ww} in the Lagrangian [Eq.~\eqref{e.Lag}] to find
\begin{equation}
L=4\pi\phi_0^2\int \dd r\,r^2\left[ -\chi\frac12f^{\prime 2}+\chi^3\frac12f^2\omega^2-\chi^3U(f)/\phi_0^2\right].
\end{equation}
The variation of the Lagrangian with respect to $\chi$ has two parts: first the explicit dependence on $\chi$, and second the 
variation of functions that now depend on $\chi$, $f(r)\to f(\chi\,r)$. This second collection of terms, with $\chi$ then 
set to 1, is 
 the usual variation of the Lagrangian, and so it vanishes by definition for solutions of the field equation. By requiring the 
other term in the variation to also vanish when $\chi=1$, we find
\begin{equation}
\frac32\frac{\omega Q}{4\pi}=\phi_0^2\int \dd r\,r^2\left[\frac12f^{\prime 2}+3U(f)/\phi_0^2\right],\label{e.Lconst}
\end{equation}
which can be used to rewrite the energy as
\begin{equation}
E=\omega Q+\frac{4\pi}{3}\phi_0^2\int \dd r\,r^2f^{\prime 2}\,.\label{e.EGlobal}
\end{equation}
This expression for $E$ is useful not only because it reduces the number of integrals to be performed, but also because 
it illustrates that 
$E=\omega Q$ up to terms that depend on the \emph{derivative} of $f(r)$. Such terms turn out to be subleading in the regime 
of large Q-balls.

\subsection{The Potential}

An explicit scalar potential which has a stable vacuum and which
admits Q-ball solutions is
\begin{align}
U(\phi)
= m_\phi^2 |\phi|^2 - \beta |\phi|^4 + \frac{\xi}{m_\phi^2} |\phi|^6 \,,
\label{eq:potential}
\end{align}
where $\beta$ and $\xi$ are positive dimensionless constants. Note that by keeping only the renormalizable terms, one can 
never achieve Q-balls in a single-field stable potential. Therefore, some non-renormalizable terms must be included in the effective 
scalar potential. The usual effective field theory expectation is that
the $|\phi|^6$ term, which could be generated by introducing additional heavy scalar 
particles, accounts for the largest of these effects. Higher-order terms would then be suppressed by additional powers of 
the heavy scalar mass scale.  
Therefore, this specific polynomial potential is expected to encapsulate the essential properties of 
many high-energy scenarios that lead to Q-balls, making it of general interest. Along with this generality, however, comes 
the fact that to date, no exact solutions to this potential have been found, making accurate approximations particularly 
valuable.

Qualitatively, the negative $|\phi|^4$ term leads to an attractive interaction between $\phi$ particles that is crucial 
for forming Q-balls, while the $|\phi|^6$ term stabilizes the potential for large $\phi$ values.
For this potential we find from Eq.~\eqref{e.Omega0}
\begin{equation}
\phi_0=m_\phi\sqrt{\frac{\beta}{\xi}}\,, \ \ \ \ \omega_0=m_\phi\sqrt{1-\frac{\beta^2}{4\xi}}\,.
\end{equation}
The condition on $\omega_0$ given in Eq.~\eqref{e.Omega0} for the existence of global Q-balls translates into 
$0<\beta^2 \leq 4\xi$.
It is convenient to rewrite the potential
in terms of $\phi_0$ and $\omega_0$, rather than $\beta$ and $\xi$, as 
\begin{equation}
U(f)/\phi_0^2=\frac12(m_\phi^2-\omega_0^2)f^2\left(1-f^2 \right)^2+\frac{\omega_0^2}{2}f^2\,.
\end{equation}
The form of the Lagrangian in Eq.~\eqref{e.Lag} suggests that the more useful quantity is $V(f)$:
\beq
\frac12\omega^2f^2-U(f)/\phi_0^2
=\left(m_\phi^2-\omega_0^2 \right)\frac12f^2\left[\kappa^2-\left(1-f^2\right)^2\right]\equiv\left(m_\phi^2-\omega_0^2 \right)
\cdot V(f)\,.\label{e.Vdef}
\eeq
We have here defined the dimensionless quantity
\beq
\kappa^2\equiv\frac{\omega^2-\omega_0^2}{m_\phi^2-\omega_0^2}\,,
\eeq
which uniquely parametrizes the scalar profiles, as shown below.
Finally, we focus on the universal aspects of the Q-ball system by
switching to the dimensionless radial coordinate $\rho$~\cite{Ioannidou:2004vr},
\beq
\rho \equiv r\sqrt{m_\phi^2-\omega_0^2 }\,,
\eeq
with $\rho \in [0,\infty)$.
This leads to the equations and expressions used in the remainder of this article.
The Lagrangian $L$, charge $Q$, and energy $E$ can be expressed in terms of dimensionless integrals via
\begin{align}
L &=\frac{4\pi\phi_0^2}{\sqrt{m_\phi^2-\omega_0^2}}\int \dd \rho\,\rho^2\left[ -\frac12f^{\prime 2}+V(f)\right],
\label{e.dimLessLag}\\
Q &=\frac{4\pi\omega\phi_0^2}{(m_\phi^2-\omega_0^2)^{3/2}}\int \dd \rho\,\rho^2f^2\,,\label{e.dimLessQ}\\
E &=\omega Q+\frac{4\pi\phi_0^2}{3\sqrt{m_\phi^2-\omega_0^2}}\int \dd\rho\,\rho^2f^{\prime 2}\,,\label{e.dimLessE}
\end{align}
where primes here and hereafter denote derivatives with respect to $\rho$.
From the Lagrangian in Eq.~\eqref{e.dimLessLag}, we obtain the differential equation that determines the dimensionless 
profile $f(\rho)$ as
\begin{align}
f''+\frac{2}{\rho}f'+\frac{\dd V}{\dd f}=0\,, \quad \text{ with } \quad V(f) = \frac12f^2\left[\kappa^2-\left(1-f^2\right)^2
\right] ,
\label{e.GlobalfEq}
\end{align}
where the boundary condition $f(\rho\to \infty)=0$ produces a localized solution.
Note that this equation is singular at $\rho=0$ unless we also impose the boundary condition $f'(0)=0$.
It remains for us to solve Eq.~\eqref{e.GlobalfEq}, which depends exclusively on the dimensionless parameter $\kappa$. As we 
shortly show, $0< \kappa <1$ ($\omega_0< \omega< m_\phi$) in order to obtain Q-ball solutions. Note that we restrict 
ourselves to finding \emph{ground-state} solutions, for which $f(\rho)$ is monotonic.

\subsection{Energy Considerations for Q-Balls}

In order to solve Eq.~\eqref{e.GlobalfEq}, it is useful to study the potential $V(f)$, illustrated in the left panel 
of Fig.~\ref{f.PotentialAndProfile}.
The extrema of $V(f)$ are at $f=0$ and
\begin{equation}
f_{\pm}^2=\frac13\left(2\pm\sqrt{1+3\kappa^2} \right) .
\end{equation}
 One finds that $f_+$ is always a maximum, while $f_-$ is a minimum for $\kappa<1$ (or $\omega<m_\phi$). 
 The center of the potential,
$f=0$, is a maximum for $\kappa<1$.
When $\kappa=1$, we find that $f_-=0$, and the potential becomes nearly flat at $f=0$.

We further our understanding of the dynamics by noting that were it not for the  
friction term, $2f'/\rho$, we could write the equation of motion in Eq.~\eqref{e.GlobalfEq} as
\begin{equation}
f''+\frac{\dd V}{\dd f}=\frac{1}{f'}\frac{\dd\,}{\dd\rho}\left(\frac12f^{\prime 2}+V(f) \right)=0~,
\end{equation}
and we identify
\begin{equation}
 \mathcal{E}=\frac12f^{\prime 2}+V(f) \label{e.EConserved}
 \end{equation}
 as the conserved energy of the system~\cite{Paccetti:2001uh};
 evaluated at $\rho=0$, we see that $\mathcal{E}=0$.
 This quantity is not conserved when the friction term
$2f'/\rho$ is included; instead we find 
\begin{equation}
\frac{\dd\mathcal{E}}{\dd\rho}=f'\left( f''+\frac{\dd V}{\dd f}\right)=-\frac{2}{\rho}f^{\prime 2}\,.
\end{equation}
 We can integrate this over the Q-ball trajectory: this starts at
$ f=f(0)$ with $f'=0$, and ends at $ f=0$ with $f'=0$.
As we have taken $V(0)=0$, we find~\cite{Mai:2012yc}
 \begin{equation}
 V(f(0))=2\int_0^\infty \dd\rho\,\frac{f^{\prime 2}}{\rho}\,.\label{e.FricWork}
 \end{equation}
Thus, we see that the difference in height of the two potential peaks (see the left panel of 
Fig.~\ref{f.PotentialAndProfile}) must be equal 
to the energy lost due to friction.

This immediately leads to a qualitative understanding of the Q-ball trajectories.
 Particle trajectories that begin near $f_+\approx0$ 
 must transition to the true vacuum without much friction. Consequently, the transition must begin when the friction term is 
suppressed by large $\rho$, after which it proceeds quickly.
On the other hand, as the energy difference between $V(f_+)$ and $V(0)=0$ increases, 
the friction cannot completely compensate for the change. 
Therefore, these trajectories start further and further below $f_+$, and so begin their transitions earlier and 
earlier, leading to smaller-radius Q-balls with softer edges. For large enough $\kappa$,
 the trajectory must start so far down the first maximum that there is very little rapid motion. The particle takes a 
long time rolling to smaller values, so the soft edge of the Q-ball extends out 
further and further until at $\kappa=1$ the trajectory begins and ends at rest at $f=0$.
This picture is confirmed in the next section using numerical solutions, shown in the right panel 
of Fig.~\ref{f.PotentialAndProfile}.

\subsection{Coleman's Thin Wall}
\label{sec.coleman}

One special trajectory is the thin-wall limit, which is defined by $\kappa\to 0$ and implies $V(0)=V(f_+)=0$. Because the maxima have equal heights, no energy 
can be lost to friction along the particle's path.
Since the friction is suppressed by $1/\rho$,  the particle trajectory starts at $f=f_+$ and remains at this value for 
infinite radius, 
after which it can roll from one maximum to the other without loss of energy.   The final motion from near the top of the maximum near $f=0$ to the maximum also takes 
infinite radius, but most of the
 transition takes place over a short range. Such solutions are called ``thin wall" because the transition region (or wall) 
is small compared to the radius of the Q-ball.
From the above, it is clear that $\kappa\to 0$ also implies $R\to \infty$. This relationship is quantified in Sec.~\ref{sec.improved_thinwall}.

In this thin-wall limit (i.e.~$\kappa\to 0$, $\omega\to \omega_0$, $R\to \infty$), Q-balls are often studied using 
Coleman's thin-wall ansatz~\cite{Coleman:1985ki}:
\begin{equation}
f(\rho)=\left\{\begin{array}{cr}
1\,, & ~~~~~~~~\rho<R^\ast\,,\\
0\,,& ~~~~~~~~~\rho> R^\ast\,,
\end{array} \right.
\label{e.thinwallprofile}
\end{equation}
where $R^\ast$  denotes the radius of the Q-ball.
For this profile we can  evaluate the integral in Eq.~\eqref{e.dimLessQ} to find the Q-ball charge
\bea
Q=\frac{4\pi}{3}\left(\frac{R^\ast}{\sqrt{m_\phi^2-\omega_0^2}}\right)^3\phi_0^2\, \omega_0\,. \label{e.globalQthin}
\eea
The Q-ball charge is hence  $\phi_0^2 \omega_0$ times the volume of the Q-ball, at least for $\omega_0 >0$. The case $\omega_0=0$ is discussed in Sec.~\ref{sec.improved_thinwall}.

We could evaluate the Q-ball energy [Eq.~\eqref{e.dimLessE}] using $f'(\rho)=0$ to obtain $E=\omega_0 Q$, but neglecting the 
singular point at $f'(R^\ast)$ is only correct as $R^\ast\to\infty$. We can do better by writing the relevant 
integral in Eq.~\eqref{e.dimLessE} as
\beq
\int_0^\infty\dd\rho\,\rho^2 f^{\prime2}= R^{\ast2}\int_0^{1}\dd f f^{\prime}=\int_0^1\dd f\left. 
\sqrt{-2V(f)}\right|_{\kappa=0}\,,
\eeq
where in the first equality we have used the fact that $f'$ is only nonzero at $\rho=R^\ast$, and in the last equality we have 
taken the thin-wall ``energy" $\mathcal{E}$ in Eq.~\eqref{e.EConserved} as conserved and zero. This leads to the following 
thin-wall expression for the energy:
\beq
E=\omega_0 Q+\frac{\pi\phi_0^2}{3\sqrt{m_\phi^2-\omega_0^2}}R^{\ast2} \,,\label{e.globalEthin}
\eeq
where the first term is interpreted as the volume contribution to the energy and the second as the surface 
contribution.\footnote{Coleman's derivation~\cite{Coleman:1985ki} leads to a surface term which is 3 times greater. 
This results from assuming the integral over the potential, and not  just the kinetic term, is dominated 
at $\rho=R^\ast$, which is not correct. }

The simple profile in Eq.~\eqref{e.thinwallprofile} characterizes Q-balls in the thin-wall limit $\kappa=0$, but says nothing about finite $\kappa$. As the numerical solutions in the next section make clear, it is a rather poor estimate of the features of the profile away from the infinite-radius limit. Unsurprisingly, the numerical values of $Q$ and $E$ also diverge drastically from the thin-wall prediction as $\kappa$ is increased.

\section{ Numerical Solutions}
\label{sec.numerics}

As suggested in the previous section, it is useful to make an analogy between the field
profiles and one-dimensional particle motion~\cite{Coleman:1985ki}.
Consider a particle satisfying the equation
\beq
\ddot{x}+\frac{2}{t}\dot{x}+\frac{\dd V}{\dd x}=0\,,
\eeq
 where dots denote time derivatives. This is a particle
moving in a potential $V(x)$ experiencing time-dependent friction.
The condition that $f'(0)=0$ is analogous to $\dot{x}(0)=0$, meaning that the particle starts at rest, while the 
condition $f(\infty)=0$ corresponds to $x(\infty)=0$, which means the particle must end up at the local maximum at $x=0$. 
Thus, the profiles for $f$ correspond to the trajectories of a particle whose friction decreases with time, rolling down 
a potential and ending up at the top of a local peak. It is often convenient to use this language, familiar from 
mechanics, to describe 
the $f$ profiles. 

For instance, this language makes clear the range of $\kappa$. For $\kappa \geq 1 $, $f=0$ is not a maximum, so the 
particle can only stop there if it begins there at rest, which is a trivial solution, or if it oscillates about the 
minimum at $f=0$, which would imply periods with $f<0$. On the other hand, for $\kappa < 0 $, $f=0$ is higher
than any other maxima, and so nothing can roll onto it. If $\kappa=0$, then the only point on the other hill which is not 
lower than $f=0$ is exactly at the maximum $f=f_+$, and the particle never rolls from this equilibrium point.
Hence, we must have $0<\kappa<1$, or, equivalently, $\omega_0<\omega<m_\phi$.

\begin{figure}[t]
\centering
\includegraphics[width=0.49\textwidth]{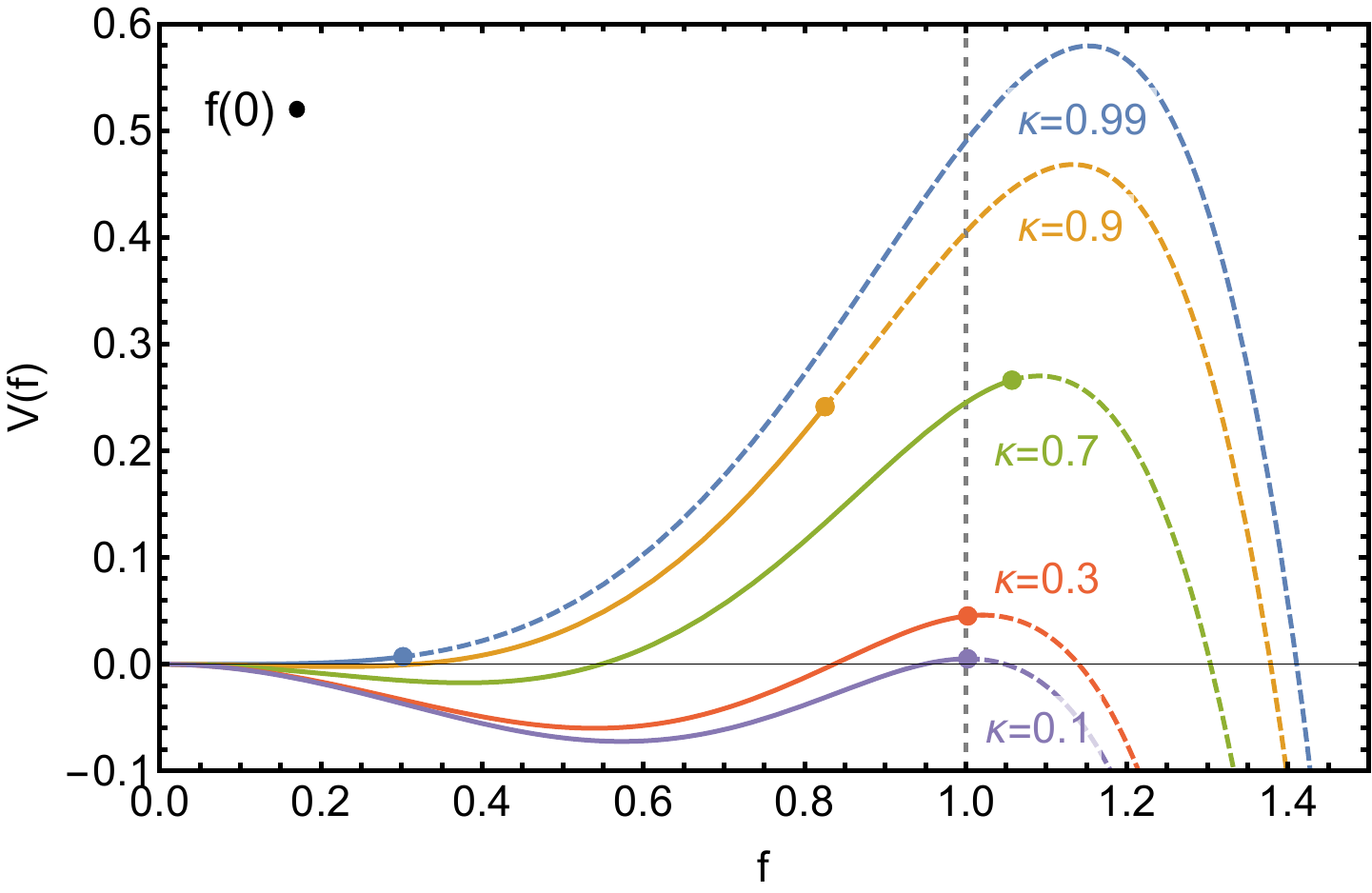}
\includegraphics[width=0.47\textwidth]{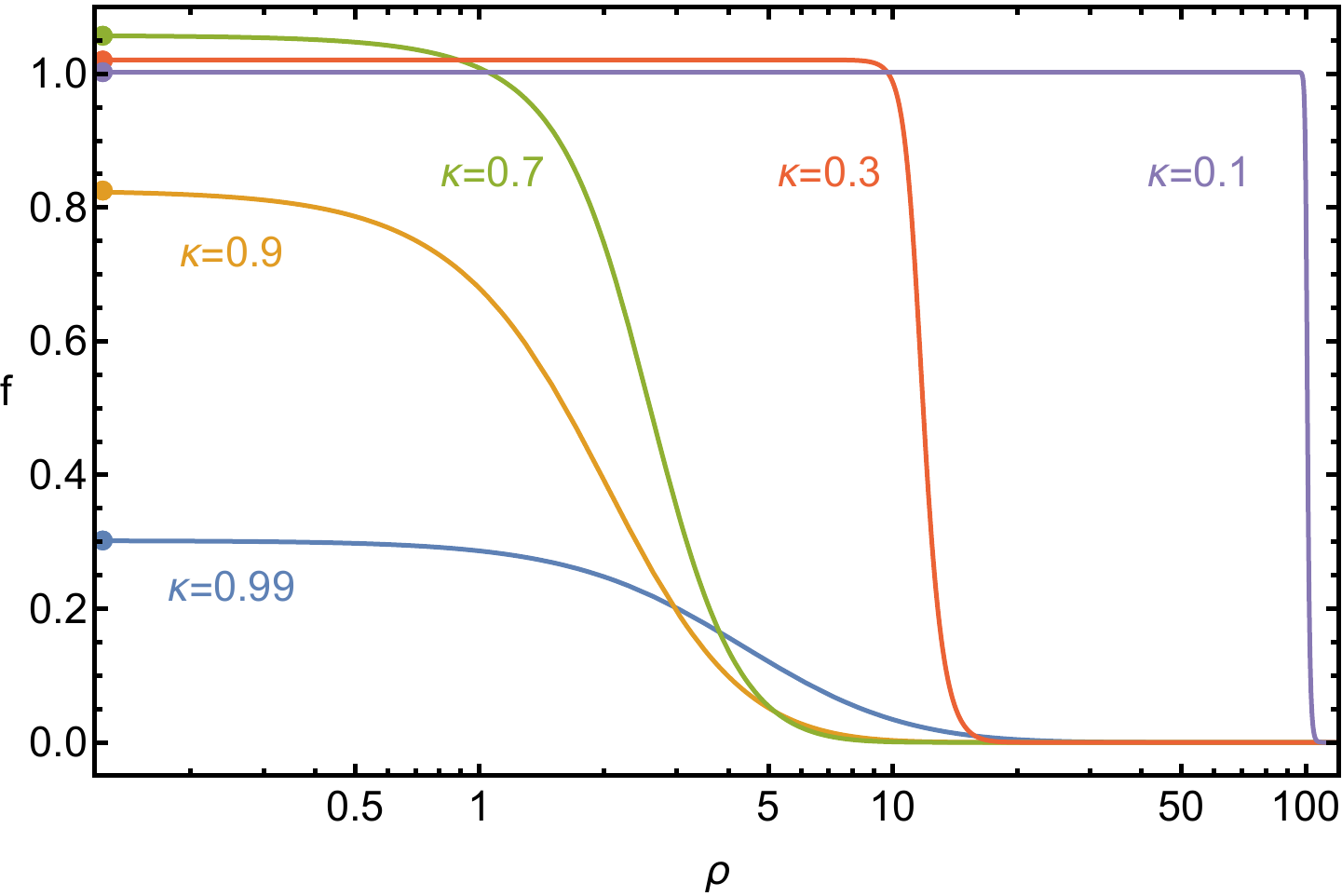}
\caption{
Plot of the effective Q-ball potential (left) and corresponding Q-ball profiles (right) for several values of $\kappa$. 
The solid curves of the potential denote the trajectory of the scalar field as it rolls from rest at the point $f(0)$. The 
vertical dashed 
line on the left plot denotes the location of the maximum when $\kappa=0$.
}
\label{f.PotentialAndProfile}
\end{figure}

 Notice, however, that while the true thin-wall limit is singular,  
the profile for $\kappa=0.1$ already demonstrates qualitatively similar behavior to Coleman's profile, Eq.~\eqref{e.thinwallprofile}, in Fig.~\ref{f.PotentialAndProfile}. The 
trajectory begins near $f_+$ and remains there for a large radius. Then it rolls very quickly, and without much friction,  
to the top of the other maximum.

The rolling particle language also suggests the \emph{shooting method} for solving the $f$ equation numerically~\cite{Coleman:1985ki}. To employ 
this technique, one 
specifies an initial value 
for $f$ at $\rho=0$ and numerically integrates the equation out to 
$\rho\gg1$ to determine whether $f$ makes it up to $f=0$ or rolls too far. The initial value of $f$ is then adjusted 
until the field comes 
to a stop on the $f=0$ point of the effective potential. This shooting method is easy to implement numerically and can 
be used to 
generate Q-ball solutions. 

One can also interpret the differential equation [Eq.~\eqref{e.GlobalfEq}] as a vacuum tunneling process. Several computer codes 
have been written specifically to solve these equations efficiently, and can also be used for Q-balls. One example 
is \texttt{AnyBubble}~\cite{Masoumi:2016wot}, which we have used to check our results.

An alternative method we have used is to solve the boundary value problem directly. In order to enforce the 
boundary condition at $\rho=\infty$ we change variables to
\beq
y=\frac{\rho}{1+\rho/a}\,,
\eeq
where $a$ is some positive number. This maps the range $\rho\in[0,\infty)$ to $y\in[0,a]$. The differential 
equation~\eqref{e.GlobalfEq} becomes
\beq
\left(1 - \frac{y}{a}\right)^4\left( \frac{\dd^2f}{\dd y^2} +\frac{2}{y} \frac{\dd f}{\dd y}\right)+\frac{\dd V}{\dd f}=0\,,
\eeq
with the boundary conditions $\frac{\dd f}{\dd y}(0)=0$ and $f(a)=0$. Given a sufficiently accurate guess for the profile $f(y)$,  
standard finite-element methods quickly converge to the exact solution without resorting to the shooting method's tedious 
fine-tuning of initial values. 
In the sections that follow, we obtain analytical results that act as good guides to this numerical method.

In Fig.~\ref{f.PotentialAndProfile}, we plot both the potentials and 
profiles for several values of $0<\kappa<1$. In the left plot, the actual single-particle trajectories that lead to the 
Q-ball solutions are plotted as solid lines, while the remainder of the potential is dashed. The initial position of 
the particle, corresponding to the value of $f(0)$ at the center of the Q-ball, is marked by a solid point. The 
corresponding $f(\rho)$ profiles are shown in the right plot, and are analogous to the position of the particle as a 
function of time.
The right plot in Fig.~\ref{f.PotentialAndProfile} shows that Q-balls become bigger and sharper edged for smaller 
$\kappa$, with a profile that strongly resembles a step function. These are appropriately denoted as \emph{thin-wall} 
profiles. Conversely, when $\kappa \to 1$, the Q-balls become more fuzzy and approach the trivial vacuum solution $f=0$. As 
shown below, these \emph{thick-wall} Q-balls are unstable.

Figure~\ref{f.PotentialAndProfile} also makes clear that one can define a \emph{radius} for the Q-balls because they are 
localized solutions. There is 
some ambiguity in defining the radius; here we choose the point $\rho=R^\ast$ where $f''(R^\ast)=0$.
The dimensionless quantity $R^\ast$ is related to the true dimensionful radius by
\beq
R=\frac{R^\ast}{\sqrt{m_\phi^2-\omega_0^2}}\,.
\eeq
Since the profile $f$ is fully determined by the parameter $\kappa$, there must be a relation between $R^\ast$ and $\kappa$. Indeed, 
we find below that the profile is defined more naturally in terms of $R^\ast$. Consequently, finding the relation between 
$R^\ast$ and $\kappa$ to better accuracy than previous results in the literature is essential to an accurate characterization 
of the Q-balls.


\section{An Improved Thin-Wall Profile}
\label{sec.improved_thinwall}
This section outlines our new analytic results. First, we take Coleman's thin-wall ansatz and extend it away from the $\kappa=0$ limit. We find that this already reveals the leading relation between $\kappa$ and $R^\ast$. This gives this simple approximation much more predictive power and begins to approximate the numerical results. We then focus on accurately describing the true Q-ball profile analytically.

To obtain a better profile, we divide up the space into three qualitatively distinct regions. These are the
 Q-ball interior and exterior, which correspond to $\rho \ll R^*$ and $\rho \gg R^*$,  respectively, and the surface or 
edge of the Q-ball  at $\rho \sim R^*$, which describes the transition between the interior and exterior.  Of these, the surface profile has historically been the most difficult to analyze. We find an exact result for the surface profile in the large-$R^\ast$ limit, which is a strikingly accurate fit to the entire Q-ball profile near the thin-wall limit. The profiles for the regions are then joined together, and  $R^\ast$, $Q$, and $E$ are determined.

\subsection{Advancing Beyond The Thin-Wall Limit}

While the thin-wall $Q$ and $E$ derived in Sec.~\ref{sec.coleman} were obtained for $\omega=\omega_0$, one could hope that they are approximately true away from that limit. 
We generalize Eqs.~\eqref{e.globalQthin} and~\eqref{e.globalEthin} to $\omega\gtrsim \omega_0$ via
\begin{align}
Q=\frac{4\pi}{3}\left(\frac{R^\ast}{\sqrt{m_\phi^2-\omega_0^2}}\right)^3\phi_0^2\, \omega\,, \ \ \ \ E=\omega Q+\frac{\pi\phi_0^2}{3\sqrt{m_\phi^2-\omega_0^2}}R^{\ast2}\,,\label{e.QEaway}
\end{align}
but we cannot compare them to the numerical results without knowing how $R^\ast$ depends on $\omega$ (or equivalently on $\kappa$).
However, by combining these results with the exact relation derived from Eq.~\eqref{e.dEdw},
\beq
\frac{\dd E}{\dd R^\ast}=\omega(R^\ast)\frac{\dd Q}{\dd R^\ast}\,,\label{e.dEdQ}
\eeq
we can determine how $R^\ast$ and $\kappa$ are related as we move away from the thin-wall limit. We evaluate the left-hand side 
of Eq.~\eqref{e.dEdQ} to find
\begin{align}
\frac{\dd E}{\dd R^\ast} &= \omega\frac{\dd Q}{\dd R^\ast}+Q\frac{\dd \omega}{\dd R^\ast}+\frac{2\pi\phi_0^2}
{3\sqrt{m_\phi^2-\omega_0^2}}R^{\ast}\nonumber\\
&= \omega\frac{\dd Q}{\dd R^\ast}+\frac{4\pi}{3}\left(\frac{R^\ast}{\sqrt{m_\phi^2-\omega_0^2}}\right)^3\phi_0^2
\omega\frac{\dd \omega}{\dd R^\ast}+\frac{2\pi\phi_0^2}{3\sqrt{m_\phi^2-\omega_0^2}}R^{\ast} \,.
\end{align}
This implies the following differential equation for $\omega$:
\beq
\omega\frac{\dd \omega}{\dd R^\ast}=-\frac{m_\phi^2-\omega_0^2}{2R^{\ast2}}\,,
\eeq
which can be integrated from the thin-wall limit, with $\omega=\omega_0$ and $R^\ast=\infty$, to general $\omega$ and 
$R^\ast$ to find
\beq
R^\ast=\frac{m_\phi^2-\omega_0^2}{\omega^2-\omega_0^2}=\frac{1}{\kappa^2}\,.\label{e.leadingRkap}
\eeq

This result gives the leading relation between $\kappa^2$, which determines the potential, and $R^\ast$, which characterizes 
the Q-ball size. Equation~\eqref{e.leadingRkap} is exactly satisfied for large $R^\ast$ or small $\kappa$ but remains an 
excellent approximation away from the thin-wall limit; even for $\kappa$ up to the limit of Q-ball stability, the deviations 
between this analytical estimate and the true numerical relation are only about $10\%$. However, these differences 
compound when applied to predicting $Q$ and $E$ as functions of $\omega$ by inserting Eq.~\eqref{e.leadingRkap} into Eq.~\eqref{e.QEaway}. The resulting deviations from the numerical 
results can be as large as 50\% for stable Q-balls. Consequently, an improved prediction of Q-ball properties 
relies on a better understanding of how $\kappa$ and $R^\ast$ are related. 
In general, we would have an expansion
\beq
\kappa^2(R^\ast)=\frac{1}{R^\ast}+\frac{\delta}{R^{\ast2}}+\cdots~,\label{e.kappOfR}
\eeq
where successive coefficients in the expansion can be obtained by more accurately describing the Q-ball profile.

For $\omega_0 >0$, the above expressions for $Q$, $E$, and $R$ in the thin-wall limit imply the  scaling 
$Q\propto E\propto R^3$ one expects for a lump of Q-matter~\cite{Coleman:1985ki}. For $\omega_0=0$, on the 
other hand, $\omega \propto 1/\sqrt{R}$, and thus $Q\propto R^{5/2}$ and $E\propto R^{2}$~\cite{Paccetti:2001uh}. This 
illustrates that even though the scalar profile $f$ only depends on the parameter $\kappa$, the physical Q-ball properties 
must be discussed in terms of the original Lagrangian parameters and, in particular, depend on $\omega_0$.

\subsection{ Exterior}

We now begin to more carefully determine the Q-ball profile by considering the exterior. In this case, we do not assume the thin-wall limit $\kappa\to 0$, but we keep 
$\kappa$ general. For all Q-balls, $f$ is near the true vacuum $f\sim 0$ when $\rho\gg R^\ast$. We can then approximate 
the potential as a quadratic and find the differential equation for the exterior:
\begin{align}
0 &= f''+\frac{2}{\rho}f'+\left.\frac{\dd V}{\dd f}\right|_{f=0}+f\left.\frac{\dd^2V}{\dd f^2}\right|_{f=0}+\cdots\nonumber\\
 &\simeq f''+\frac{2}{\rho}f'-(1-\kappa^2)f \,.
\end{align}
In the last equation, we have dropped terms of order $f^3$ and higher because $f\ll 1$.
 Enforcing the boundary condition $f(\infty)=0$, one finds (as obtained in Ref.~\cite{Lee:1988ag}) the exterior 
solution $f_>$ is
\beq
f_>=\frac{c_>}{\rho}e^{-\sqrt{1-\kappa^2}\rho}\,,\label{e.fGreater}
\eeq
with $c_>$ an integration constant. This exponential drop-off behavior applies to all Q-balls, but in the thin-wall regime,  
one can further set $\kappa\to 0$ in $f_>$.

\subsection{ Interior}

Near the thin-wall limit, the value of $f$ within the Q-ball is close to the maximum of the potential, $f\sim f_+$. This again 
allows us to approximate the potential as a quadratic, leading to the $f$ equation for the interior:
\begin{align}
0 &={f}''+\frac{2}{\rho}{f}'
+\left.\frac{\dd V}{\dd f}\right|_{f=f_+}+(f-f_+)\left.\frac{\dd^2V}{\dd f^2}\right|_{f=f_+}+\cdots \\
&\simeq {f}''+\frac{2}{\rho}{f}'-\alpha^2(f-f_+)\,,
\end{align}
where the neglected terms are of higher order in $(f-f_+)$, and we have defined
\beq
\alpha^2 \equiv \frac43\left( 1+3\kappa^2+2\sqrt{1+3\kappa^2}\right) 
= 4 + 8\kappa^2 - 3 \kappa^4 +\mathcal{O}(\kappa^6)\,.
\eeq
After enforcing the boundary condition ${f}'(0)=0$, one finds (as obtained in Ref.~\cite{Paccetti:2001uh}) the solution 
$f_<$ in the Q-ball interior 
\beq
f_<=f_++c_<\frac{\sinh(\alpha\rho)}{\rho}\,,
\label{interiorsol}
\eeq
where $c_<$ is an integration constant. In the thin-wall regime, where $\kappa$ is small, we can take $\alpha=2$ in $f_<$.

\subsection{Transition Region}

We now turn to the region which joins the interior and exterior.
Rather than expanding the 
potential around an extremum, we find a limit in which the dynamics can be solved exactly. In the surface region of 
the Q-ball, $\rho \sim R^*$, it is useful to use the coordinate $z=\rho-R^\ast$ and focus on the region around $z=0$. 
As the surface profile $f_s$ describes the transition from one potential maximum to the other, we must consider the 
full potential. We can write the differential equation~\eqref{e.GlobalfEq} as 
\bea
\frac{\dd^2 f_s}{\dd z^2}+\frac{2}{R^\ast+z}\frac{\dd f_s}{\dd z}+f_s\left[\kappa^2-(1-f_s^2)(1-3f_s^2) 
\right] = 0 \,.\label{e.TransEq}
\eea
The second term is suppressed by $(R^\ast)^{-1}$ and can be neglected
when $-z\ll R^*$.
More precisely, one can expand the profile as a power series in $(R^\ast)^{-1}$:
\beq
f_s(z)=f_s^{(0)}(z)+(R^\ast)^{-1}f_s^{(1)}(z)+\cdots~,\label{e.fSeries}
\eeq
and find that the leading-order profile satisfies the equation
\beq
\frac{\dd^2 f_s^{(0)}}{\dd z^2}=f_s^{(0)}\left[1-(f_s^{(0)})^2 \right]\left[ 1-3 (f_s^{(0)})^2\right].
\label{e.leadingordertransition}
\eeq
Here we have used the leading-order relation $\kappa^2=(R^\ast)^{-1}$ from Eq.~\eqref{e.kappOfR}.

Because the friction term is absent from this leading equation~\eqref{e.leadingordertransition}, the quantity 
$\mathcal{E}$ of Eq.~\eqref{e.EConserved} is conserved. 
Furthermore, as $f'(0)=V(0)=0$ 
for Q-ball solutions, we see that $\mathcal{E}=0$.
This leads to the first-order differential equation
\beq
\frac{\dd f_s^{(0)}}{\dd z}=\pm f_s^{(0)}\left(1-f_s^{(0)2} \right) ,\label{e.f0Eq}
\eeq
which is equivalent to the second-order equation given in Eq.~\eqref{e.leadingordertransition} but can be directly integrated 
to find
\beq
f_s^{(0)}(z)=\left[1+c_s e^{\pm2z} \right]^{-1/2} ,
\eeq
where $c_s$ is an integration constant.
Because $f_s$ is a monotonically decreasing function, we must take the positive sign in the exponent. 

The constant $c_s$ is determined by
requiring $f''(R^\ast)=0$, thus properly identifying $R^\ast$ with the Q-ball radius. This yields the final form of the 
transition function at leading order in $1/R^\ast$:
\beq
f_s^{(0)}(\rho)=\left[1+2 e^{2(\rho-R^\ast)} \right]^{-1/2} .\label{transition}
\eeq
This profile was considered in Ref.~\cite{Ioannidou:2003xn} in relation to Q-balls from a sixth-order potential, 
but with a different interpretation.
Despite being derived in the limit of $\rho$ close to $R^\ast$, we show below that this profile is remarkably close to the 
exact profile for all $\rho$ as long as $R^\ast$ is large. Its main shortcoming is its failure to satisfy the 
boundary condition $f'(0)=0$ away from the limit $R^\ast\to \infty$. Still, this transition profile by 
itself turns out to be an excellent approximation to the exact profile over most of the range of $\rho$,
far beyond its expected region of validity. 

Further corrections in $(R^\ast)^{-1}$ beyond the profile found above can, in principle, be obtained by solving the 
higher-order equations resulting from inserting the profile expansion given in Eq.~\eqref{e.fSeries} 
into Eq.~\eqref{e.TransEq}. 
In practice, it is difficult to obtain simple analytical results in this way.

\subsection{ Full Profile}
\label{fullansatz}

Having derived approximate expressions for the profile $f$ in the interior, exterior, and surface region, we can join them 
together to obtain the full profile. The coefficients $c_<$ and $c_>$ of the interior and exterior solutions are determined 
by enforcing continuity of $f$ and $f'$ at the matching points. 
Since the surface solution was explicitly derived in the large-$R^\ast$ limit, the expected validity of the full profile is also 
restricted to $\kappa \ll 1$.

We modify our ansatz slightly in order to improve our profile away from $\kappa\sim 0$.\footnote{We can find additional 
improvements by introducing more parameters into the profile and minimizing the resulting $E$ for a fixed $Q$, although 
this is not attempted here.} In the thin-wall 
limit, the field starts at $f(0)=f_+=1$ and transitions to $f=0$ at large $\rho$. Away from this limit, we postulate that 
the field
transitions from near the new maximum $f_+$ to zero. A natural ansatz for the 
field profile is obtained by simply rescaling the transition profile by $f_+=\frac13\left(2+\sqrt{1+3\kappa^2} \right)$.  We therefore assume that the profile takes the form
\begin{align}
f(\rho)=f_+
\begin{cases}
\displaystyle1-c_<\frac{\sinh(\alpha \rho)}{\rho} & \text{for }\rho<\rho_< \,,\\[0.25cm]
\displaystyle\left[1+2e^{2(\rho-R^\ast)} \right]^{-1/2} & \text{for }\rho_<<\rho<\rho_> \,,\\[0.25cm]
\displaystyle\frac{c_>}{\rho}e^{-\rho\sqrt{1-\kappa^2}} & \text{for } \rho_> <\rho \,,
\end{cases} \label{e.finalprofile}
\end{align}
where $c_{<,>}$ and $\rho_{<,>}$ are determined by requiring $f$ and $f'$ to be continuous at $\rho_{<,>}$. The details of 
this process and the resulting formulas for $c_{<,>}$ and $\rho_{<,>}$ are given in Appendix~\ref{app.matching}. 

Before comparing this profile [Eq.~\eqref{e.finalprofile}] to the numerical solutions, let us use it to refine the relationship 
between $R^\ast$ and $\kappa^2$ by using the energy requirement given in Eq.~\eqref{e.FricWork}:
\beq
V(f(0))=2\int_0^\infty\frac{\dd \rho}{\rho}(f^{\prime})^2\,.
\eeq
This integral is evaluated in each region separately, but one quickly finds that the interior and exterior regions give 
contributions suppressed by at least $e^{-R^\ast}$, so we neglect them in the large-$R^\ast$ limit of interest 
here. In Appendix~\ref{app.matching}, we find the leading-order results
\beq
f(0)\approx f_+\,, \ \ \ \ \rho_<\approx \frac{R^\ast}{2}\,, \ \ \ \ \rho_>\approx 2R^\ast\,,
\eeq
so we can rewrite the energy-lost-to-friction relation as
\begin{align}
V(f_+)=8f_+^2\int_{R^\ast/2}^{2R^\ast}\frac{\dd \rho}{\rho}\frac{e^{4(\rho-R^\ast)}}{\left[1+2e^{2(\rho-R^\ast)} 
\right]^3}=8f_+^2\int_{-R^\ast/2}^{R^\ast}\frac{\dd z}{z+R^\ast}\frac{e^{4z}}{\left[1+2e^{2z} \right]^3} \,,
\end{align}
where we have changed variables to $z=\rho-R^\ast$ in the last equality. Because the integrand is sharply peaked at 
$z=0$, we can consistently expand $(z+R^\ast)^{-1}$ in powers of $z/R^\ast$ and extend the limits of integration 
out to infinity, up to exponentially suppressed terms. This leads to the relation
\beq
R^\ast(\kappa)=\frac{f_+^2}{2V(f_+)}~
= \frac{1}{\kappa^2} + \frac14-\frac{5 \kappa^2}{16} + \mathcal{O}(\kappa^4)\,,
\label{e.radius}
\eeq
which agrees with Eq.~\eqref{e.leadingRkap} to lowest order in $\kappa^2$ 
but contains subleading corrections that improve the agreement with the numerical solutions.

 \begin{figure}[t]
\centering
\includegraphics[width=0.9\textwidth]{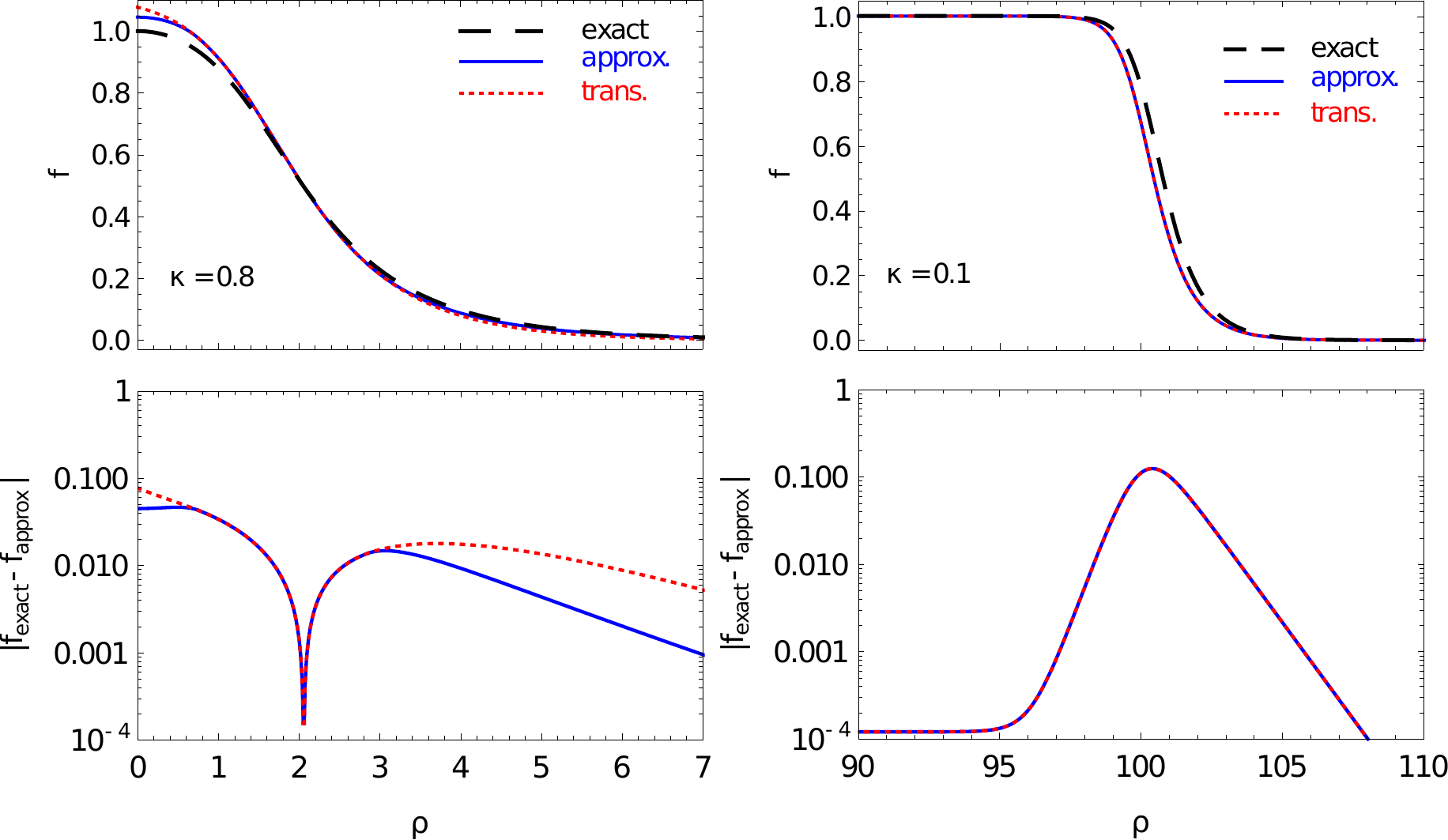}
\caption{
Top: the profile $f(\rho)$ for $\kappa=0.8$ (left) and $\kappa=0.1$ (right).
In black dashed lines, we show the exact numerical profile, in blue lines the profile of Eq.~\eqref{e.finalprofile}, and in dotted 
red lines the transition profile $f_s(\rho)$  extended beyond its region of validity. Note that within the bounds 
of the right-hand plot, the transition profile and the profile of Eq.~\eqref{e.finalprofile} coincide. 
Bottom: the difference between the exact profile and the two approximations.
}
\label{f.ProfileCompare}
\end{figure}

In Fig.~\ref{f.ProfileCompare}, we compare our full profile~\eqref{e.finalprofile} [containing the improved 
$R^\ast (\kappa)$ from Eq.~\eqref{e.radius}] to the full 
numerical solution. As expected, the agreement when $\kappa\ll 1$ is superb, save for a small 
mismatch of the radii. In fact, replacing our $R^\ast(\kappa)$ with the true numerical radius results in a 
per-mille level agreement in the large-$R^\ast$ limit (e.g.~$\kappa=0.1$), which illustrates how accurately our result captures the 
shape of the true profile and how important it is to determine $R^\ast(\kappa)$ more precisely. Remarkably, there 
is still good agreement between the two profiles even for rather large $\kappa$. 
While the simple step-like thin-wall ansatz from Eq.~\eqref{e.thinwallprofile} would describe the profile of 
$\kappa=0.1$ moderately well, it is a bad fit for $\kappa=0.8$, highlighting the need for a better analytical 
description. Our profile fares exceptionally well even for such large $\kappa$, despite being formally derived in 
the large-$R^\ast$ limit.

It is also striking that the transition profile $f_s$ itself provides a surprisingly good approximation to the 
full profile. In Fig.~\ref{f.ProfileCompare}, we show
$f_+\left[1+2e^{2(\rho-R^\ast)} \right]^{-1/2}$ (red dotted lines), with $\rho$ now running over the entire region 
$0\leq \rho<\infty$. For $\kappa\ll 1$, this simplified profile is indistinguishable from the full 
profile [Eq.~\eqref{e.finalprofile}], 
while the differences are rather small even for large $\kappa$. The full profile provides a better description, of 
course, but at the price of a more complicated analytical expression.
The main shortcoming of the pure transition profile is its behavior at $\rho=0$, where $f'(0)= -2 \exp (-2 R^\ast)
[1+\mathcal{O}(1/R^\ast)]$. The boundary condition $f'(0)=0$ is thus only satisfied asymptotically as 
$R^\ast\to \infty$. Still, for many practical purposes it suffices to use the simple transition profile 
together with Eq.~\eqref{e.radius}.

\subsection{Charge and Energy}

Using the improved profile from Eq.~\eqref{e.finalprofile}, we can calculate the charge $Q$ and energy $E$ from
Eqs.~\eqref{e.dimLessQ} and~\eqref{e.dimLessE}, respectively. 
The integrals of interest are $\int \dd \rho \, \rho^2 f^2$ and $\int \dd \rho \, \rho^2 (f')^2$, which can be performed 
analytically, although the expressions are  long and largely unenlightening. For large $R^\ast$, they read
\begin{align}
\begin{split}
\int \dd \rho \, \rho^2 f^2 &\simeq \frac{f_+^2 R^{\ast 3}}{3}\left(  1 - \frac{3\ln 2}{2 R^\ast} + 
\frac{\pi^2 + 3\ln^2 2}{4 R^{\ast 2}} -\frac{(\pi^2+\ln^22)\ln2}{8R^{\ast3}}\right) ,\\
\int \dd \rho \, \rho^2 f'^2 &\simeq  \frac{f_+^2 R^{\ast 2}}{4}\left(  1+\frac{1-\ln2}{R^\ast}+
\frac{\pi^2+(\ln2 -2)3\ln2}{12R^{\ast2}} \right) .
\end{split}
\label{e.integrals}
\end{align}
Together with Eq.~\eqref{e.radius}, we can also obtain an expansion in small $\kappa$.
Notice that the first integral is directly proportional to the dimensionful Q-ball volume 
$ 4\pi (m_\phi^2 - \omega_0^2)^{-3/2} \int \dd \rho \, \rho^2 f^2 $; in the large-$R$ limit we then 
find the volume is $\simeq 4\pi R^3/3$, which shows that our definition of the radius via $f''(R^\ast)=0$ is 
particularly sensible in the thin-wall limit.

Using Eqs.~\eqref{e.dimLessQ} and \eqref{e.dimLessE}, we can then obtain our final expressions for $Q$ and $E$, expanded 
again in large $R^\ast$, because this is the expected regime of validity of our underlying scalar profile:
 \begin{align}
 \begin{split}
 Q &\simeq \frac{4\pi}{3}\left(\frac{R^\ast}{\sqrt{m_\phi^2-\omega_0^2}} \right)^3\omega\phi_0^2
f_+^2\left[1-\frac{3\ln2}{2R^\ast}+\frac{\pi^2+3\ln^22}{4R^{\ast2}}-\frac{(\pi^2+\ln^22)\ln2}{8R^{\ast3}} \right] ,\\
 E &\simeq \omega Q+\frac{\pi\phi_0^2f_+^2}{3\sqrt{m_\phi^2-\omega_0^2}}R^{\ast2}\left[ 1+\frac{1-\ln2}{R^\ast}+
\frac{\pi^2+(\ln2 -2)3\ln2}{12R^{\ast2}}\right] .
 \end{split}
 \label{e.QandE}
 \end{align}
Together with Eq.~\eqref{e.radius}, we now have an analytical approximation for $Q$ and $E$ as a function of the potential 
parameters. Other quantities, such as pressure, can be calculated straightforwardly~\cite{Mai:2012yc}. 
These expressions become exact in the limit $\omega\to \omega_0$ ($\kappa\to 0$, $R^\ast\to \infty$), where they agree 
with Coleman's thin-wall result. However, our expressions are also approximately valid for smaller $R^\ast$, as 
shown in the next section where they are compared to numerical results, courtesy of the subleading $1/R^\ast$ terms.

Let us first discuss the theoretical validity of our expressions for $Q$ and $E$. As shown in Eq.~\eqref{e.dEdw}, global 
Q-ball solutions must fulfill the relation $\dd E/\dd \omega = \omega \dd Q/\dd \omega$. 
Expressing $\omega$ as a function of the radius by inverting Eq.~\eqref{e.radius}, we can check this relation order by order 
in $1/R^\ast$ with our expressions from Eq.~\eqref{e.QandE} and find that $\dd E/\dd R^\ast = 
\omega (R^\ast) \dd Q/\dd R^\ast$ is valid 
up to terms of order $(R^\ast)^0$. One could imagine taking the $Q$ and $E$ results as exact, and then 
require that Eq.~\eqref{e.dEdw} be satisfied to determine $\kappa(R^\ast)$. The result is $\kappa^2 = 1/R^\ast + 
(1+\ln 16)/(8 R^{\ast\,2})+\mathcal{O}(1/R^{\ast\,3})$, but as the improvement with respect to Eq.~\eqref{e.radius} 
is marginal, we do not pursue this further here.

We have yet to address Q-ball stability~\cite{Friedberg:1976me,GRILLAKIS1987160,Lee:1991ax,Tsumagari:2008bv}, 
mainly because it has no bearing on solving the differential equation~\eqref{e.GlobalfEq}. Once interpreted in a 
physical context, however, stability further restricts the allowed values of $\kappa$. While there are several 
criteria for Q-ball stability, the strongest one in our case is also the one that is easiest to understand. This 
is the criterion that the energy of a stable Q-ball must be less than the mass of $Q$ free scalars,
\begin{align}
E< m_\phi Q \,.
\label{e.stability}
\end{align}
If Eq.~\eqref{e.stability} is not satisfied, the Q-ball can decay.
Using our approximations, we can re-express this stability requirement as
\begin{align}
\frac{\omega}{m_\phi} \lesssim \frac{1}{14} \left(5+\sqrt{41}\right)  + \left(\frac{1}{8}+\frac{37}{56 \sqrt{41}}\right) 
\frac{\omega_0^2}{m_\phi^2}
\end{align}
for $\omega_0\ll m_\phi$. 
Notice that stability does not depend purely on $\kappa$ but also on $\omega_0/m_\phi$. Using numerical results, we can show 
that this dependence is rather weak and that the region of stability is between $\kappa\lesssim 0.82$ (for $\omega_0=0$) 
and $\kappa\lesssim 0.84$ (for $\omega_0\sim m_\phi$).\footnote{The analysis of stability may be more subtle when 
$\omega_0=0$, see Ref.~\cite{Paccetti:2001uh}.}
From Fig.~\ref{f.integrals}, we can already see that our analytic approximations are valid precisely in the stable 
Q-ball regime, while failing to describe the unstable thick Q-balls that arise for $\kappa \gtrsim 0.84$.\footnote{It is worth emphasizing that the regions of stability can change significantly with the type of potential that produces the Q-balls~\cite{Mai:2012yc}. For instance, in Ref.~\cite{Kusenko:1997ad} is is shown that the Q-balls which spring from a potential with a cubic term can be stable even for $\omega\to m_\phi$.}

\section{ Comparison Between Numerics and Analytics }
\label{sec.comparison}

 \begin{figure}[t]
\centering
\includegraphics[width=0.49\textwidth]{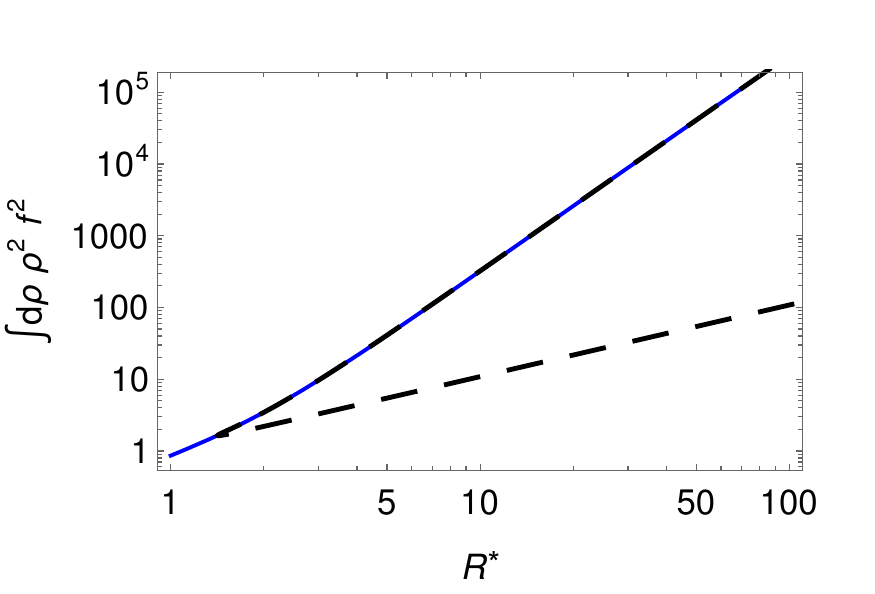}
\includegraphics[width=0.49\textwidth]{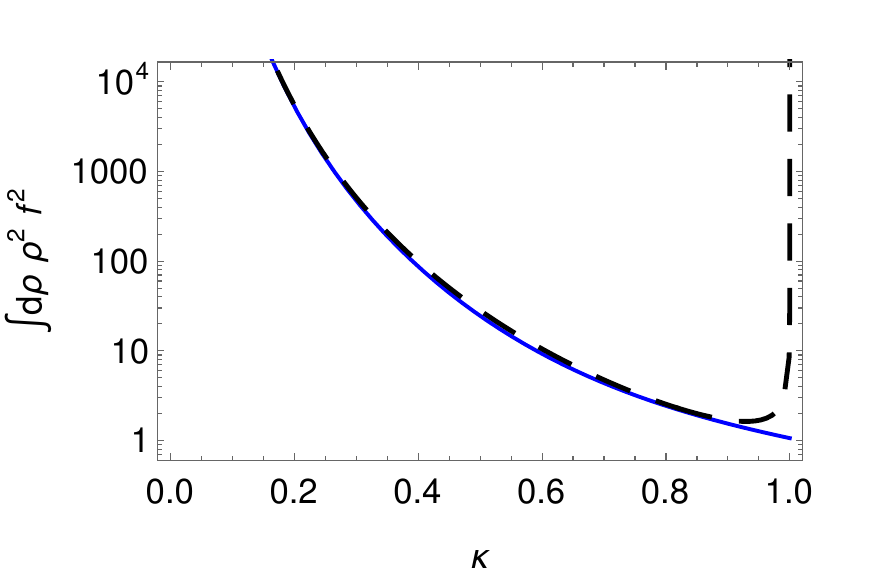}\\
\includegraphics[width=0.49\textwidth]{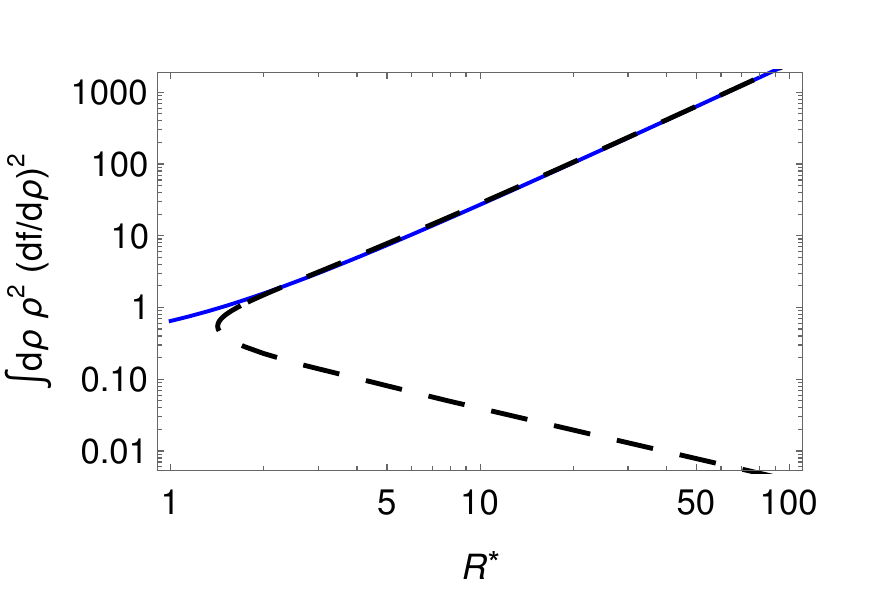}
\includegraphics[width=0.49\textwidth]{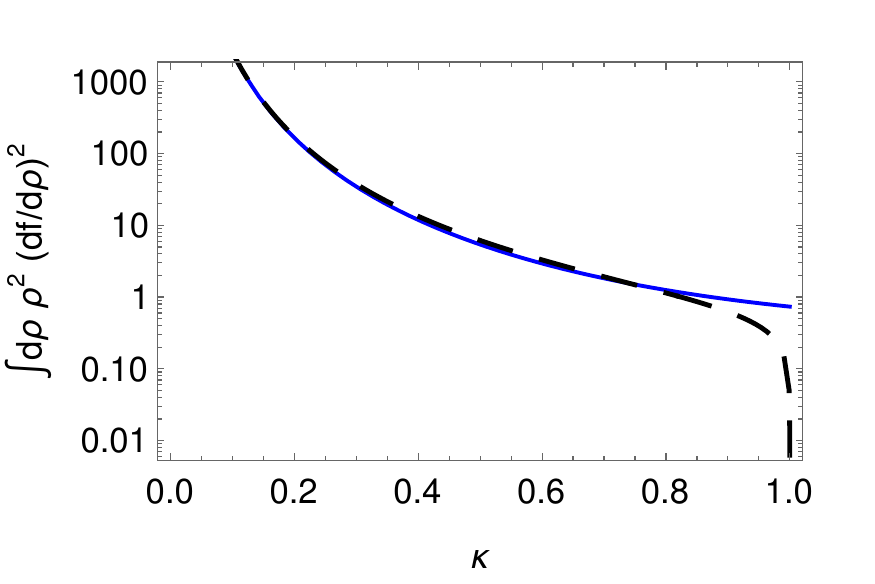}
\caption{The integrals $\int \dd \rho \, \rho^2 f^2$ (top) and $\int \dd \rho \, \rho^2 f'^2$ (bottom) as a function of 
$R^\ast$ (left) and $\kappa$ (right). The dashed black line denotes the exact numerical solution and the blue 
solid line our prediction from Eq.~\eqref{e.integrals}.}
\label{f.integrals}
\end{figure}

In this section, we test our analytic results by comparing them to numerical solutions. We start by comparing the two 
integrals of Eq.~\eqref{e.integrals}.
These analytical expressions are shown along with the exact numerical values in Fig.~\ref{f.integrals}. As a function of 
$R^\ast$, we see a remarkable agreement between our prediction and one branch of the exact numerical solution. As a 
function of $\kappa$, our expressions successfully approximate the numerical solution for $\kappa \lesssim 0.8$. This is 
precisely the regime where Q-balls are stable, and hence most interesting for most physical applications.
Again, as a function of $R^\ast$, the analytic approximation of $\int \dd \rho \, \rho^2 f^2$ agrees with the numerical 
results to better than $1\%$ for stable Q-balls; as a function of $\kappa$, this agreement worsens to up to $13\%$ (near 
$\kappa\sim 0.5$) due to our imperfect $R^\ast(\kappa)$ modeling.
The integral $\int \dd \rho \, \rho^2 (f')^2$ agrees with the numerical results to better than $5\%$ for $R^\ast > 2$ and 
to better than $13\%$ for $\kappa \lesssim 0.8$; again, our formula for $R^\ast(\kappa)$ introduces a rather large error.

\begin{figure}[t]
\includegraphics[width=1\textwidth]{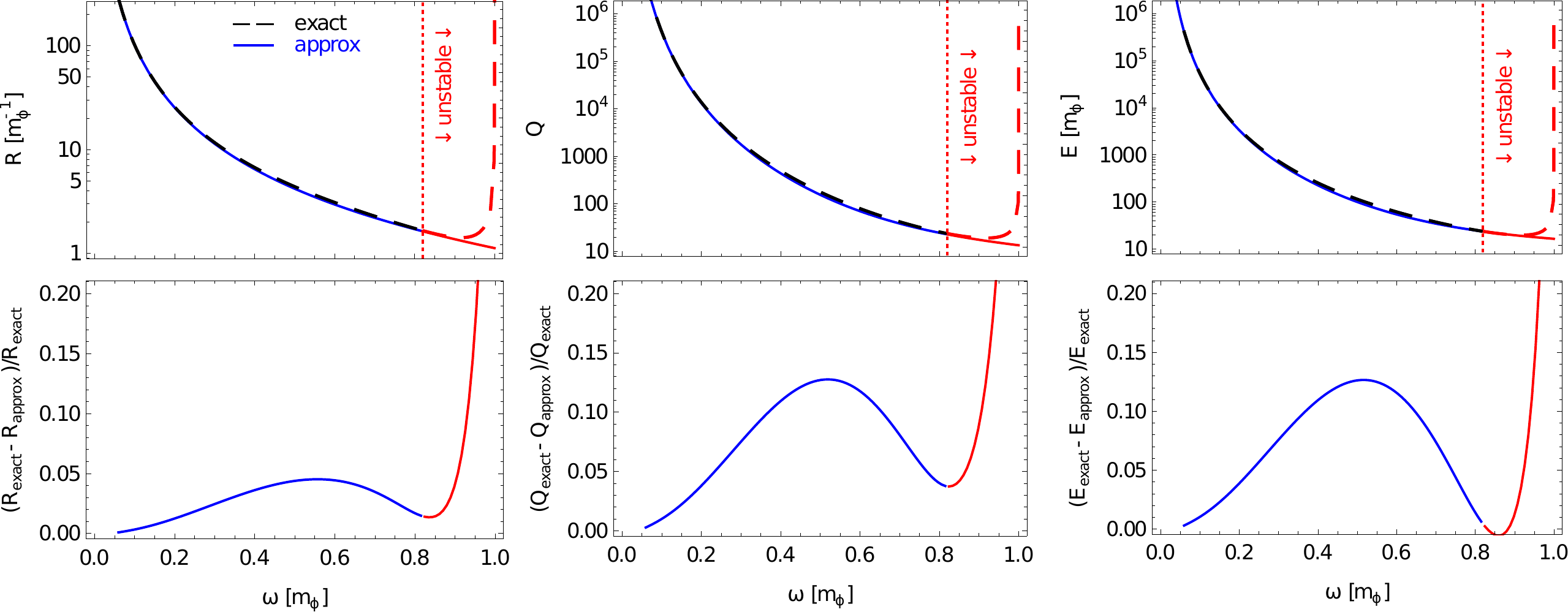}
\caption{
$R(\omega)$ (left), $Q(\omega)$ (middle), and $E(\omega)$ (right) for the potential parameter set $\phi_0=m_\phi$, 
$\omega_0=0$. In the upper row, black/red dashed lines show the exact numerical values in the stable/instable regime, and 
the blue and red lines show our approximation of Eq.~\eqref{e.QandE}. In the lower row, we show the relative difference 
between exact and approximate values.
The red marked region $\omega \gtrsim 0.82$ indicates unstable Q-balls with $E > m_\phi Q$.
}
\label{fig:comparison}
\end{figure}

In Fig.~\ref{fig:comparison} (left), we compare our analytic approximation of $R^\ast(\omega)$ from Eq.~\eqref{e.radius} to 
numerical results with $\phi_0=m_\phi$, $\omega_0=0$. 
Beyond $\omega\sim 0.82 m_\phi$, the Q-balls have $E > m_\phi Q$
and are unstable to fission. Our analytical estimate in Eq.~\eqref{e.radius} agrees with the numerical results to better 
than $5\%$ in the stable Q-ball regime!
As expected, in the thin-wall limit $\omega\to\omega_0$, the
agreement becomes exact, since our profile [Eq.~\eqref{e.finalprofile}] approaches the exact profile.

In Fig.~\ref{fig:comparison}, we also show $Q$ (middle) and $E$ (right) as functions of $\omega$. Again, the numerical
and analytical results are in excellent agreement -- better than 13\% -- over the entire stable region. Recall that the 
agreement was only to 50\% using the simple formulas in Eq.~\eqref{e.QEaway}. As in that case, small deviations in our $R^\ast(\omega)$ from the true relationship clearly 
propagate into larger mismatches in $Q$ and $E$, which could be alleviated by using an 
improved $R^\ast(\omega)$.

\begin{figure}[t]
\includegraphics[width=0.49\textwidth]{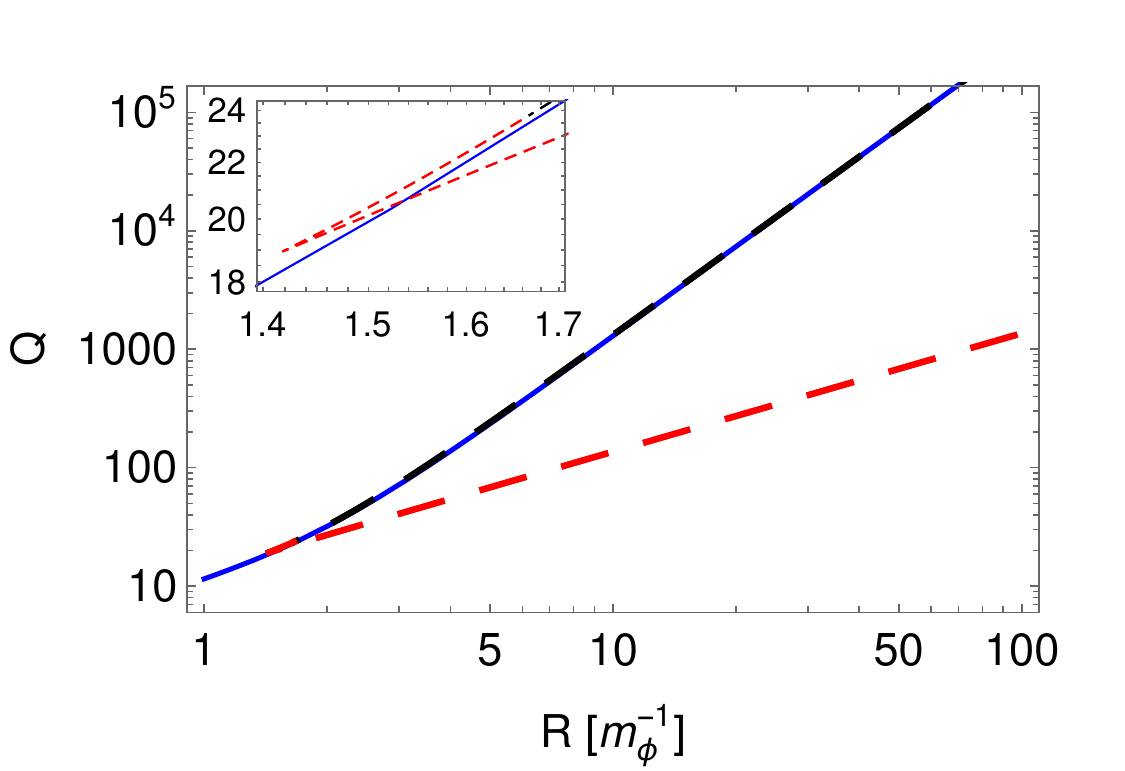}
\includegraphics[width=0.49\textwidth]{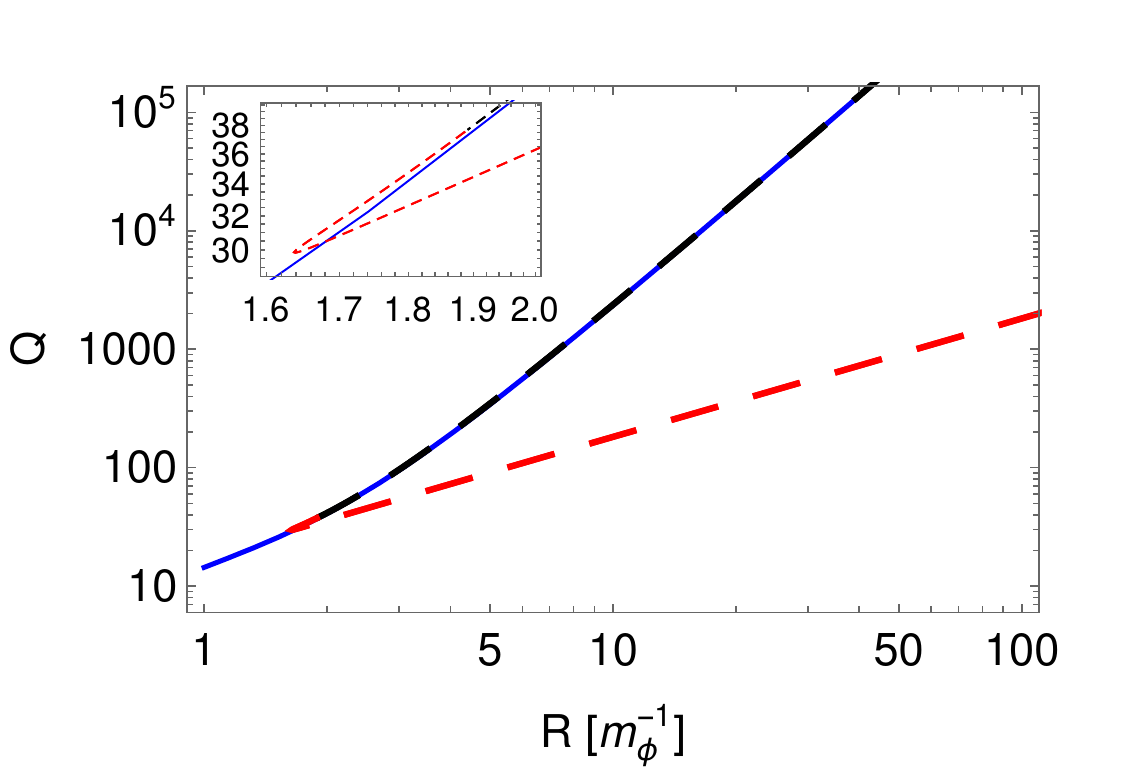}\\
\includegraphics[width=0.49\textwidth]{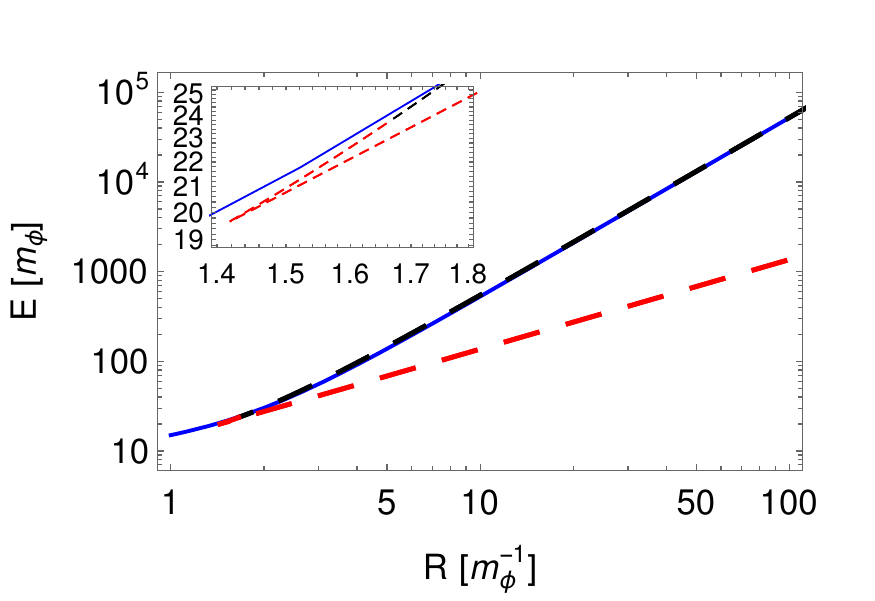}
\includegraphics[width=0.49\textwidth]{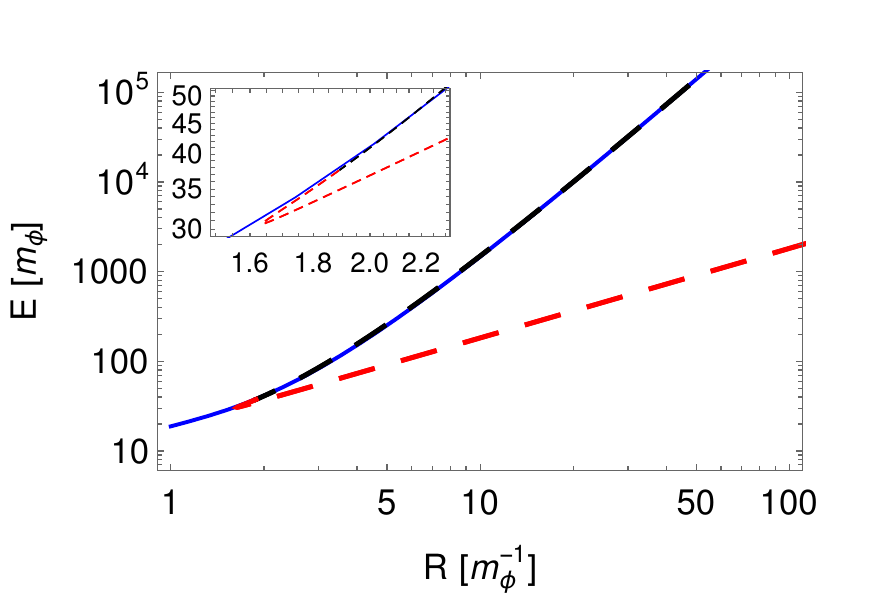}
\caption{
$Q$ vs.~$R$ for $\phi_0=m_\phi$, $\omega_0=0$ (left), $\omega_0=m_\phi/2$ (right). The dashed black line is the exact solution in the 
stable regime; the dashed red line is unstable. The blue solid line is our prediction from Eq.~\eqref{e.QandE}. 
}
\label{fig:QandEvR}
\end{figure}

Finally, by inverting Eq.~\eqref{e.radius} to obtain $\omega (R^\ast)$, we can obtain $Q(R^\ast)$ and $E(R^\ast)$, which 
are compared to the numerical results in Fig.~\ref{fig:QandEvR}. 
The agreement is superior to what is shown in Fig.~\ref{fig:comparison} -- better than 3\% for $Q(R^\ast)$ and 5\% for 
$E(R^\ast)$ for stable Q-balls. Again, the simple results in Eq.~\eqref{e.QEaway} led to agreement as poor as 40\%, so an order of 
magnitude in improvement has been achieved. This again demonstrates that $R^\ast(\omega)$ is the dominant source of error, while 
the explicit dependence on the  radius in Eq.~\eqref{e.QandE} is very accurate.

In Fig.~\ref{fig:QandEvR}, we also show the qualitative difference of the two cases $\omega_0=0$ (left) and $\omega_0>0$ (right).
For large $R$ and $\omega_0\neq 0$, $E\propto Q\propto R^3$, as expected from the thin-wall approximation. 
However, as 
observed in Ref.~\cite{Paccetti:2001uh}, for $\omega_0=0$, $\omega\sim \sqrt{m_\phi/R}$ for large $R$ and 
thus $Q\propto R^{5/2}$ and $E\propto R^2$. In the unstable regime, not covered by our approximations, we find 
numerically that 
$E\simeq m_\phi Q\propto R$ for large $R$ for all $\omega_0$.

\section{Conclusion}
\label{sec.conclusion}

We have provided a guide to understand Q-ball solutions and provided simple formulae for  describing their 
salient characteristics which can be used without numerically solving the underlying differential equations. Our analytical approximations 
significantly improve upon Coleman's well-known thin-wall solution both qualitatively and quantitatively.
We can describe stable Q-balls in arbitrary $U(1)$-symmetric sextic potentials to about $10\%$ accuracy, 
and much better for larger Q-balls. 

Our expectation from effective field theory is that these potentials capture the leading dynamics which produce Q-balls, and 
are therefore of particular interest. Consequently, we expect our results to be useful for simply and accurately finding the 
properties of  Q-balls for use in various cosmological and astrophysical studies. In addition, our approach has been general, 
allowing the complete set of Q-ball-producing sextic potentials to be modeled and studied, even though no exact solutions for 
this potential are known.

Furthermore, we have derived a scalar profile that closely describes the exact solution to the differential equation 
over a wide range of parameters. We have also obtained accurate analytical formulae for the resulting Q-ball 
properties, namely the charge, energy, and radius. The procedure employed here should be useful in finding analytic 
approximations for Q-balls in other space-time dimensions or potentials and, more generally, to other similar 
differential equations, for instance, in the context of vacuum decay.
Finally, our improved description of global Q-balls paves the way to a better description of the significantly 
more difficult \emph{gauged} Q-balls, which will be discussed in a separate article.

\section*{Acknowledgements}
We thank Alexander Kusenko for clarifying communications. This work was supported in part by NSF Grant No.~PHY-1915005. C.B.V.~also acknowledges support from Simons 
Investigator Award \#376204.

\appendix

\section{Matching Conditions for the Full Profile}
\label{app.matching}

In this appendix, we determine the conditions on the profile of Eq.~\eqref{e.finalprofile},
\begin{align}
f(\rho)=f_+
\begin{cases}
\displaystyle1-c_<\frac{\sinh(\alpha \rho)}{\rho} & \text{for }\rho<\rho_< \,,\\[0.25cm]
\displaystyle\left[1+2e^{2(\rho-R^\ast)} \right]^{-1/2} & \text{for }\rho_<<\rho<\rho_> \,,\\[0.25cm]
\displaystyle\frac{c_>}{\rho}e^{-\rho\sqrt{1-\kappa^2}} & \text{for } \rho_> <\rho \,,
\end{cases}
\end{align}
that make $f$ and $f'$ continuous at $\rho_{<,>}$. First, we 
find that $\rho_<$ is determined by the equation
\beq
2\rho_<=\left(\sqrt{1+2e^{2(\rho_<-R^\ast)}}-1 \right)\left(2+e^{-2(\rho_<-R^\ast)} \right)
\left[ \alpha\rho_<\coth\left( \alpha\rho_<\right)-1\right] .
\eeq
This has no simple solution, but if we assume $\rho_<\ll R^\ast$ and $\alpha\rho_<\gg1$, we find
\beq
\rho_<\approx\frac{1}{\alpha-2}\approx\frac12R^\ast\,,
\eeq
where we have used $\kappa^2=1/R^\ast$. The constant $c_<$ takes the form
\beq
c_<=\frac{2e^{2(\rho_<-R^\ast)}\rho_<^2}{\left[ \alpha\rho_<\cosh(\alpha\rho_<)-\sinh(\alpha\rho_<)\right]
\left[ 1+2e^{2(\rho_<-R^\ast)}\right]^{3/2}}
\approx R^\ast e^{-2 R^\ast}\,.
\eeq
So, we find that the interior solution joins the surface solution about halfway between the center of the 
Q-ball and the edge, and that only for smaller $R^\ast$ does the $\sinh$ term play a significant role. Note that we also 
find a prediction for the value of the profile in the center of the Q-ball:
\beq
f(0)=f_+\left(1 -\alpha c_<\right) .
\eeq
This shows how the initial value of $f$ on the potential is away from the maximum as $R^\ast$ becomes smaller, but only 
by exponentially small amounts.

Turning to $\rho_>$, we find the matching condition
\beq
e^{-2(\rho_>-R^\ast)}=2\left(\frac{\rho_>}{1+\rho_>\sqrt{1-\kappa^2}}-1 \right) ,
\eeq
which leads the approximate solution
\beq
\rho_>\approx 2R^\ast.
\eeq
We also find the integration constant $c_>$ to be
\beq
c_>=\rho_>e^{\rho_>\sqrt{1-\kappa^2}}\left[1+2e^{2(\rho_>-R^\ast)} \right]^{-1/2}
\approx \sqrt{2}R^\ast e^{R^\ast}\,.
\eeq
In this case, we find that the matching point to the exterior solution is well beyond $R^\ast$. Thus, we again find that 
much of the full profile is approximated by the transition solution.

\bibliographystyle{utcaps_mod}
\bibliography{BIB}

\end{document}